\documentclass[12pt,letterpaper,preprint]{aastex}

\usepackage{amsmath}
\usepackage{amsfonts}
\usepackage{amssymb}

\def\harmtd{{\tt HARM3D}}

\author{Hotaka Shiokawa\altaffilmark{1}, Joshua C. Dolence\altaffilmark{2}, 
Charles F. Gammie\altaffilmark{1,3}, Scott C. Noble\altaffilmark{4}}

\altaffiltext{1}{Department of Astronomy, University of Illinois at Urbana-Champaign, 1002 West Green Street, Urbana, IL 61801}

\altaffiltext{2}{Department of Astronomy, University of Illinois at Urbana-Champaign, 1002 West Green Street, Urbana, IL 61801; Current address: Department of Astrophysical Sciences, Princeton
University, Princeton, NJ 08544}

\altaffiltext{3}{Department of Physics, University of Illinois at Urbana-Champaign, 1110 West Green Street, Urbana, IL 61801}

\altaffiltext{4}{Center for Computational Relativity and Gravitation,
School of Mathematical Sciences, Rochester Institute of Technology,
Rochester, NY 14623}

\title{Global GRMHD Simulations of Black Hole Accretion Flows: a Convergence Study}

\begin{abstract}
Global, general relativistic magnetohydrodynamic (GRMHD) simulations of
nonradiative, magnetized disks are widely used to model accreting black
holes.  We have performed a convergence study of GRMHD models computed
with {\tt HARM3D}.  The models span a factor of $4$ in linear
resolution, from $96 \times 96 \times 64$ to $384 \times 384 \times
256$.  We consider three diagnostics of convergence: (1) dimensionless
shell-averaged quantities such as plasma $\beta$; (2) the azimuthal
correlation length of fluid variables; and (3) synthetic spectra of the
source including synchrotron emission, absorption, and Compton
scattering.  Shell-averaged temperature is, except for the lowest
resolution run, nearly independent of resolution; shell-averaged
plasma $\beta$ decreases steadily with resolution but shows signs of
convergence.   The azimuthal correlation lengths of density, internal
energy, and temperature decrease steadily with resolution but show signs
of convergence.  In contrast, the azimuthal correlation length of
magnetic field decreases nearly linearly with grid size.  We argue by
analogy with local models, however, that convergence should be achieved
with another factor of $2$ in resolution.  Synthetic spectra are, except
for the lowest resolution run, nearly independent of resolution.  The
convergence behavior is consistent with that of higher physical
resolution local model (``shearing box'') calculations and with the
recent nonrelativistic global convergence studies of Hawley et al. (2011).

\end{abstract}

\begin{document}

\section{Introduction}

The numerical study of black hole accretion flows has advanced
significantly in the last decade.  The advent of techniques for
numerically solving the equations of general relativistic
magnetohydrodynamics (GRMHD) has enabled self-consistent global modeling
of accretion driven by the magneto-rotational instability (MRI)
\citep{Balb91,Gamm04_MRI} onto rotating black holes.  Qualitative
aspects of these simulations are code independent \citep[e.g.][]{DeVi03,
Gamm03, Anni05}, but quantitative variations raise the question of
numerical convergence.  Recent work has shifted focus from dynamical
properties of the accretion flow to simulated observations that can
potentially constrain parameters for particular sources such as Sgr A*
\citep{GRM, Mosc10, Dext09, Dext10}, including polarized radiative
transfer \citep{Shch10}.  To assess the credibility
of these radiative models, it is necessary to assess \emph{quantitative}
convergence of the underlying GRMHD simulations.

Convergence studies of global accretion models are computationally expensive.  An
alternative is to use a local model with shearing box boundary
conditions to study the dynamics of MRI driven turbulence.   These are
simpler in the sense that there are fewer free parameters, and cheaper
in that numerical resolution can be focused on a few correlation volumes
$\sim H^3$, where $H$ is the disk scale height.  The local model has for
decades been a key theoretical tool for probing astrophysical disks
\citep[e.g.][]{GL65, GT78, NGG87} coupled to the shearing box
boundary conditions has been widely used for the study of magnetized
disks \citep[e.g.][]{ HB91, HB92, HGB95, HGB96, Ston96, Sano04,
Hirose06, From07a, From07b, Guan09, Davi10, From10, Guan11, Simo11}.

Shearing box models have been integrated (1) with or without a mean
magnetic field; (2) with or without stratification; (3) with or without
explicit dissipation; (4) with and without explicit treatment of energy
transport.  There are now dozens of shearing box studies that treat
aspects of this problem.  The only models that clearly {\em do not}
converge are unstratified, zero-net field models without explicit
dissipation \citep{From07a}.  These models have a magnetic field
correlation length that decreases proportional to the grid scale
\citep{Guan09}. But with explicit dissipation \citep{Lesu07, From10}, a
mean field \citep{HGB95, Guan09}, or stratification \citep{Davi10,
Simo11}, the models {\em do} converge.  One of the best resolved studies
is \cite{Davi10}, who convincingly demonstrate convergence of a
stratified, isothermal, zero explicit dissipation model with the {\tt
athena} code at a physical resolution of up to $128$ zones per scale
height $H$.  These stratified local models are physically closest to
global simulations \citep[e.g.][]{DHK2}, which are
dominated by toroidal magnetic field.  Local studies have shown,
therefore, that with sufficient resolution numerical studies of
MRI-driven turbulence can converge.

Local models can focus on a few $H^3$, while global simulations
must contain many $H^3$.  Are any of the dozen or so global disk models
\citep[e.g.][ and many others]{Bran96, Mats96, DeVi03, Gamm03, DHK1, McKi04, Gamm04, McKi06, Frag07,
 Beck08, Shaf08, Beck09, Frag09b, Frag09a, Nobl09, Nobl10,
Penn10, Beck11, Floc11, HGK11} converged?  And are synthetic observations based on
global models (e.g.\citet{Dext11, Hilb10, Mosc10, Dext09, Nobl07,
Schn06}; Dolence et al. 2011 in prep.) sensitive to resolution?  While
some authors have included limited resolution studies
\citep[e.g.][]{Shaf08, Nobl10, Penn10}, the answer is not yet clear.

The first systematic convergence test of a global black hole accretion
simulation was done by \citet[][hereafter HGK]{HGK11},  using a {\tt
zeus} type code to simulate an $H/R \approx 0.1$ disk in a
pseudo-Newtonian potential.  HKG simulate a $\pi/2$ wedge in azimuth,
varying resolution around a fiducial $256 \times 288 \times 64$ (r, z, $\phi$ in
cylindrical coordinate).  After reviewing local model simulations and
global nonrelativistic models HGK find that a minimum of $10$ cells per
vertical characteristic MRI wavelength is required for convergence
\citep[HGK's $Q_z$; e.g.][]{Sano04}, and $20$ cells per azimuthal MRI
wavelength (HGK's $Q_\phi$).  They conclude that most global simulations
to date are far from resolved, except \citet{Nobl10} which used barely
adequate poloidal resolution.  

In this paper we study the same convergence problem considered by HGK,
but (1) in relativistic MHD and (2) using slightly different
diagnostics.  We ask what resolution is required for convergence (if
convergence can be achieved), and how the global resolution requirements
are related to local models.  We are also particularly interested in
whether resolution influences the spectra calculated from the models in
the weakly radiative limit.  This {\em requires} a fully relativistic
simulation since in weakly radiative accretion flows much of the
emission arises from plasma near or even inside the innermost stable
circular orbit (ISCO) of a spinning black hole.  At these radii the
relativistic models incorporate the dynamics of the plunging region and
strong lensing effects on the radiation field.

A third contrast with HGK is that we simulate a full $2\pi$ in
azimuth rather than $\pi/2$.  Our experience suggests that there is
structure in the disk in all azimuthal fourier components, with the most
power in the $m=1$ component.  Models with small azimuthal extent have
reduced field strength and therefore require higher physical resolution
by the HGK $Q$ criteria.  

We proceed as follows.  \S 2 describes the code and initial and boundary
conditions.  \S 3 describes convergence of radial profiles of
non-dimensional variables.  \S 4 describes convergence of azimuthal
correlation lengths.  \S 5 describes convergence of simulated spectra
calculated with a Monte Carlo code.  \S 6 gives a brief summary.


\section{Simulations}

Throughout the paper, we follow the standard notation of
\citet{MTW} and set $GM=c=1$.  We consider a test fluid (no
self-gravity) in the Kerr metric with dimensionless spin
$a^*=1-2^{-4}\approx0.94$.  The governing GRMHD equations express
conservation of particle number
\begin{equation}\label{H1}
(\rho u^{\mu})_{;\mu}=0 \quad \text{,}
\end{equation}
and conservation of energy-momentum
\begin{equation}\label{H2}
{T^{\mu}}_{\nu;\mu}=0 \quad \text{,}
\end{equation}
together with the source-free Maxwell equations
\begin{equation}\label{H3}
^{*}\!{F^{\mu\nu}}_{;\nu}=0 \quad \text{,}
\end{equation}
where $u^{\mu}$, $\rho$, $T^{\mu\nu}$, and $^{*}\!F^{\mu\nu}$ are the
fluid's four velocity, rest mass density, GRMHD stress-energy tensor,
and dual of the electromagnetic field tensor, respectively.  The
equation of state is
\begin{equation}
p = (\gamma - 1) u
\end{equation}
where $\gamma=13/9$, appropriate for a collisionless plasma with
relativistic electrons and non-relativistic protons.

We evolve the GRMHD equations using the \harmtd\ code
\citep{Nobl09,Nobl06,Gamm03}.  \harmtd\ is a conservative high-resolution
shock-capturing scheme demonstrated to have second order convergence in
space and time for smooth flows.  The zone-centered magnetic field is
updated with flux-interpolated constrained transport
\citep{Gamm03,Toth00} which preserves a particular numerical
representation of $\nabla\cdot\mathbf{B}=0$. For this study, we use
piecewise parabolic interpolation for both fluxes and EMFs. 

The numerical grid is uniform in modified Kerr-Schild coordinates $x_1$,
$x_2$, and $x_3$ \citep{Gamm03}:
\begin{align}
x_1 &= \ln r \\
\theta &= \pi x_2 + h \sin(2\pi x_2) \\
x_3 &= \phi
\end{align}
where $r$, $\theta$, and $\phi$ are the Kerr-Schild radius, colatitude,
and azimuth, respectively. We set $h = 0.35$ to
concentrate the grid near the equatorial plane.  The grid extends from
below the horizon to $r = 40$, [0.017$\pi$, 0.983$\pi$] in colatitude,
and [0, $2\pi$) in azimuth.  \harmtd\ sets a ``floor" for density and
internal energy to avoid numerical problems that arise when those values
are low: $\rho_{min}=10^{-4}r^{-3/2}$ and $u_{min}=10^{-6}r^{-5/2}$.

\begin{figure}
\plotone{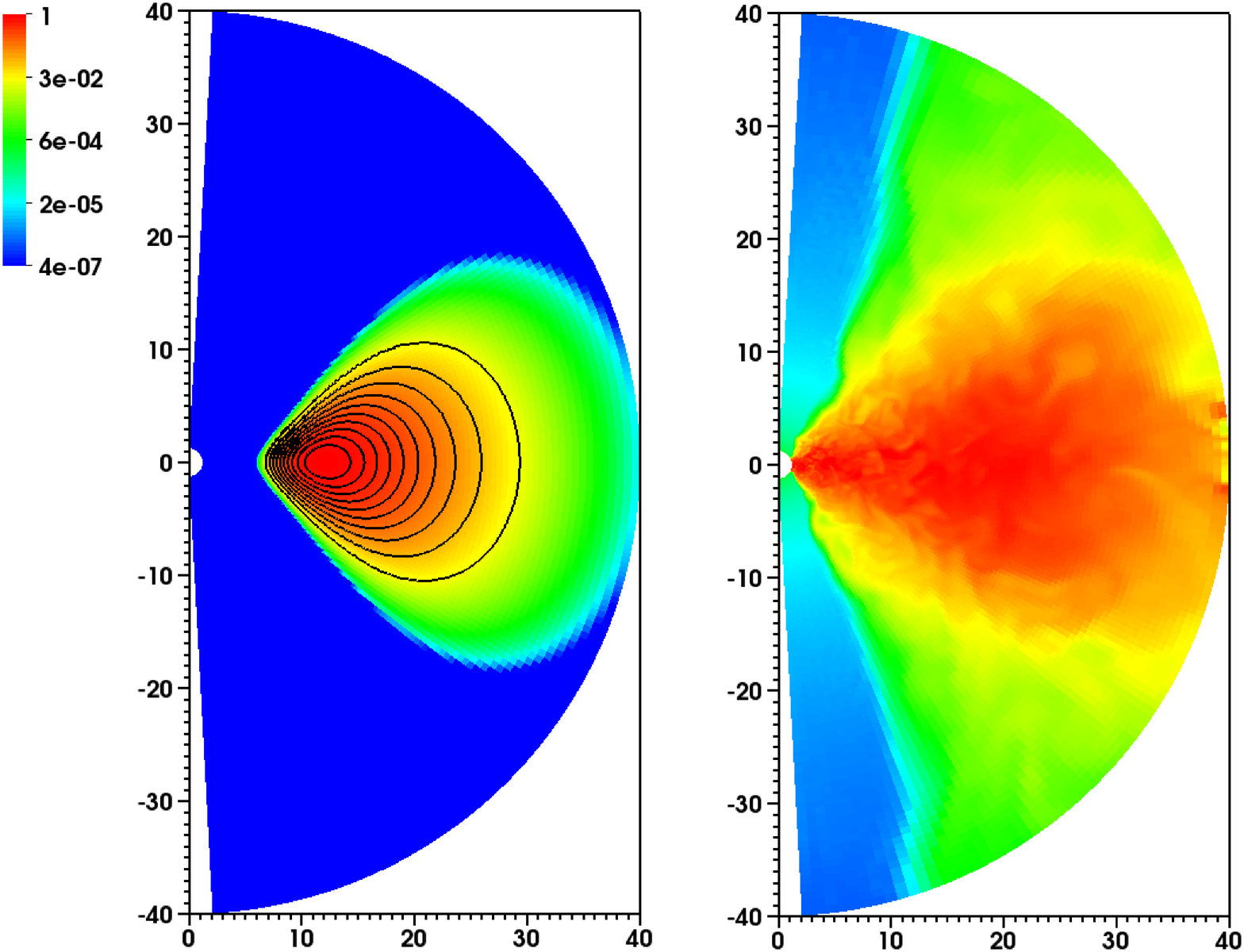} 
\caption{Poloidal slices of the initial and turbulent state of the global simulation.
The pseudo-color is showing scaled logarithmic density and black lines are the initial magnetic field lines.}
\label{snap}
\end{figure}

The initial condition is an equilibrium, prograde torus \citep{Fish76} with inner edge at $r = 6$, pressure maximum at $12$, and outer edge at $40$.
To make the torus unstable to MRI, it is seeded with weak poloidal magnetic field whose vector potential is
\begin{equation}
A_{\phi} =
	\begin{cases}
		C (\rho/\rho_{max} - 0.2)& \text{if $A_{\phi} > 0$ } \\
		0& \text{if $A_{\phi} \le$ 0}
	\end{cases}
\end{equation}
where C is a constant and $\rho_{max}$ is the maximum initial density.
This gives dipole field line loops that run parallel to density
contours.  The field strength is normalized so that the ratio of the maximum
gas pressure to maximum magnetic pressure $\beta$ is $100$.  Small
perturbations are introduced into the initial conditions to seed the
MRI.
The density and magnetic field lines are shown in Figure \ref{snap} for the initial conditions and for a later
snapshot of the turbulent accretion flow.

The models have outflow boundary conditions at the inner and outer
radial ($x_1$) boundaries and periodic boundary conditions in the
azimuthal ($x_3$) direction.  The remaining ($x_2$) boundaries are
offset slightly from the pole, so the grid excludes a narrow cone around
each pole.  This avoids having the last polar zone control the timestep
via the Courant condition because the polar zones become narrow in $x_3$
(the computational expense is proportional to $N_{x}^5$ if poles are
included!).  While this treatment is essential for a convergence study,
it is difficult to implement an appropriate boundary condition on the
cone.  We consider two different polar boundary conditions.  

The first, ``hard'' boundary is a solid reflective wall.  We manually
set the flux through the boundary to zero, and adjust the EMF in the
flux-ct routine to make the cutout completely opaque to the magnetic
field, since the field vectors are modified in the routine after setting
the boundary condition.  This boundary condition produces an unphysical
relativistic flow in the grids along the polar cone, so in addition we
force the poloidal velocity in the zones along the boundary to be zero.

The second, ``soft" boundary also models a reflective wall.  The
variables in the ghost zones are all copied from the first physical
zone.  The $x_2$ components of the velocity and magnetic field are
inverted across the boundary (as usual for reflecting boundaries), but
this only zeros fluxes on the boundary to within truncation error.
This version of the polar boundary condition permits some leakage of
magnetic flux through the polar boundaries, but does not produce
unphysical flows along the boundary.

We ran a low resolution simulation with no polar cutout to evaluate both
boundary conditions.  The results suggest that the difference between
the boundary conditions does affect the evolution of the high latitude
``funnel'' region.  The soft boundary condition, in particular, causes a
steady drop in the funnel region magnetic flux.  On the other hand, all
three cases (hard, soft, and no cutout) exhibit remarkably similar disk
evolution.

\begin{center}
{\textsc{Table 1. List of Runs}}\\
\vspace{0.3cm}
\begin{tabular}{cll}

\hline\hline
Resolution	& Duration ($\frac{GM_{\textrm{BH}}}{c^3}$)	& Polar Boundary Type\\
\hline

$96 \times 96 \times 64$	& 16,000					& Soft\\
$128 \times 128 \times 96$	& 12,000					& Soft\\
$192 \times 192 \times 128$	& 10,000					& Soft\\
$384 \times 384 \times 256$	& $\,\,6,000$					& Soft\\
$96 \times 96 \times 64$	& 16,000					& Hard\\
$128 \times 128 \times 96$	& 12,000					& Hard\\
$192 \times 192 \times 128$	& 10,000					& Hard\\

\end{tabular}
\end{center}

Our runs have numerical resolution $(N_{x_1}, N_{x_2}, N_{x_3})=$
(96, 96, 64), (144, 144, 96), (192, 192, 128), and (384, 384, 256).  The
runs last until $t_f = 16,000$ for $96 \times 96 \times 64$, $12,000$ for $144 \times 144 \times 96$,
$10,000$ for $192 \times 192 \times 128$, and $6,000$ for $384 \times 384 \times 256$.  Each resolution
is run for both the soft and hard polar-boundary conditions {\em except}
the highest resolution case which is run only for the soft-polar
boundary due to numerical expense.  A list of runs is shown in Table 1.
The runs required $\approx 10^6 (N_{x_1}/384)^4 (t_f/6,000)$ cpu hours
on TACC {\tt ranger}.

Each simulation's initial data contains noise inserted in each zone with
a random number generator.  This noise seeds the growth of instabilities
in the torus.  Each run will therefore differ in the details of the
evolution, but over long enough periods one expects the differences to
average away.  Nevertheless, because our runs have finite duration, we
expect some ``cosmic variance,'' and this noise from run-to-run
variations is present in every measurement we use to evaluate
convergence.

To evaluate run-to-run variation, we have repeated each of the $N_{x_1} =
96$ and $N_{x_1} = 144$ runs 3 times, and have used the variance of these
runs to attach error bars to our measurements.  We find that large
run-to-run variations are caused by ``events'' that last a non-negligible
fraction of the simulation time.  For example, the lowest resolution
runs sometimes gather a large mass of plasma near the ISCO, then dumps
it suddenly into the black hole.  We have also observed a bundle of
magnetic field directed opposite to the field in the funnel merge into
the funnel, leading to a large fluctuation in the run with resolution
$144 \times 144 \times 96$ and hard-polar-boundary.  While the nature, frequency, and
origin of these events is still unclear (we have only a handful of runs)
it appears that run-to-run variation decreases at higher resolution.


\section{Radial profiles of non-dimensional variables}

We will compare poloidally, azimuthally, and time averaged radial
profiles of the flow variables for the different resolution runs.  We
take a density-weighted average to focus on the accretion flow within $\sim\,H$ of the
equatorial plane.  The explicit expression for the averaged radial
profile $F(x_1)$ for a variable $f$ is
\begin{equation}\label{Ana1}
F(x_1) = \frac{\int_{t_1}^{t_2} \bar{f}(t,x_1) dt}{t_2-t_1}
\end{equation}
where
\begin{equation}\label{Ana2}
\bar{f}(t,x_1) = \frac{ \int_{(x_2)_1}^{(x_2)_2} \int_{(x_3)_1}^{(x_3)_2} \sqrt{-g} \rho(t,\vec{x}) f(t,\vec{x}) dx_2 dx_3 } { \int_{(x_2)_1}^{(x_2)_2} \int_{(x_3)_1}^{(x_3)_2} \sqrt{-g} \rho(t,\vec{x}) dx_2 dx_3 }
\end{equation}
is the density weighted poloidally and azimuthally averaged radial
profile of the variable $f$ and $g=g(\vec{x})$ is the determinant of the
metric.  For our case, $((x_2)_1,(x_2)_2)=(0.01,0.99)$ and
$((x_3)_1,(x_3)_2)=(0,2\pi)$.

We compare only non-dimensional variables since dimensional variables
depend on the accretion rate, which decreases in time as the initial
torus is accreted by the black hole.  Our choice of the non-dimensional
variables are scaled electron temperature $\theta_e = k T_e/m_e$ ($= m_p
p_g/(2 m_e \rho$) if $T_e = T_p$), and $\beta \equiv p_g / p_B = (\Gamma
-1)u/(b^2/2)$, where $b^2 \equiv b^{\mu}b_{\mu}$,
\begin{align}
b^{\mu} &\equiv \frac{1}{\gamma}({g^{\mu}}_{\nu}+u^{\mu}u_{\nu})\mathcal{B}^{\nu} \\
\mathcal{B}^{\mu} &\equiv -n_{\nu}{^* \! F}^{\mu\nu} \; \text{where} \; n_{\mu}=(-\sqrt{-1/g^{tt}},0,0,0) \; \text{,}
\end{align}
$\gamma$ is the Lorentz factor of the flow measured in the normal
observer's frame, and $\Gamma$, $k$, $m_p$ and $m_e$, and $T_p$ and
$T_e$ are the adiabatic index, Boltzmann constant, proton and electron
mass, and proton and electron temperature, respectively.  When
calculating $\beta$ we average $p_g$ and $p_b$ separately using equation
\ref{Ana2} and take the ratio of the averages.  This prevents zones with
near-zero magnetic energy from dominating the average.

\begin{figure}
\plotone{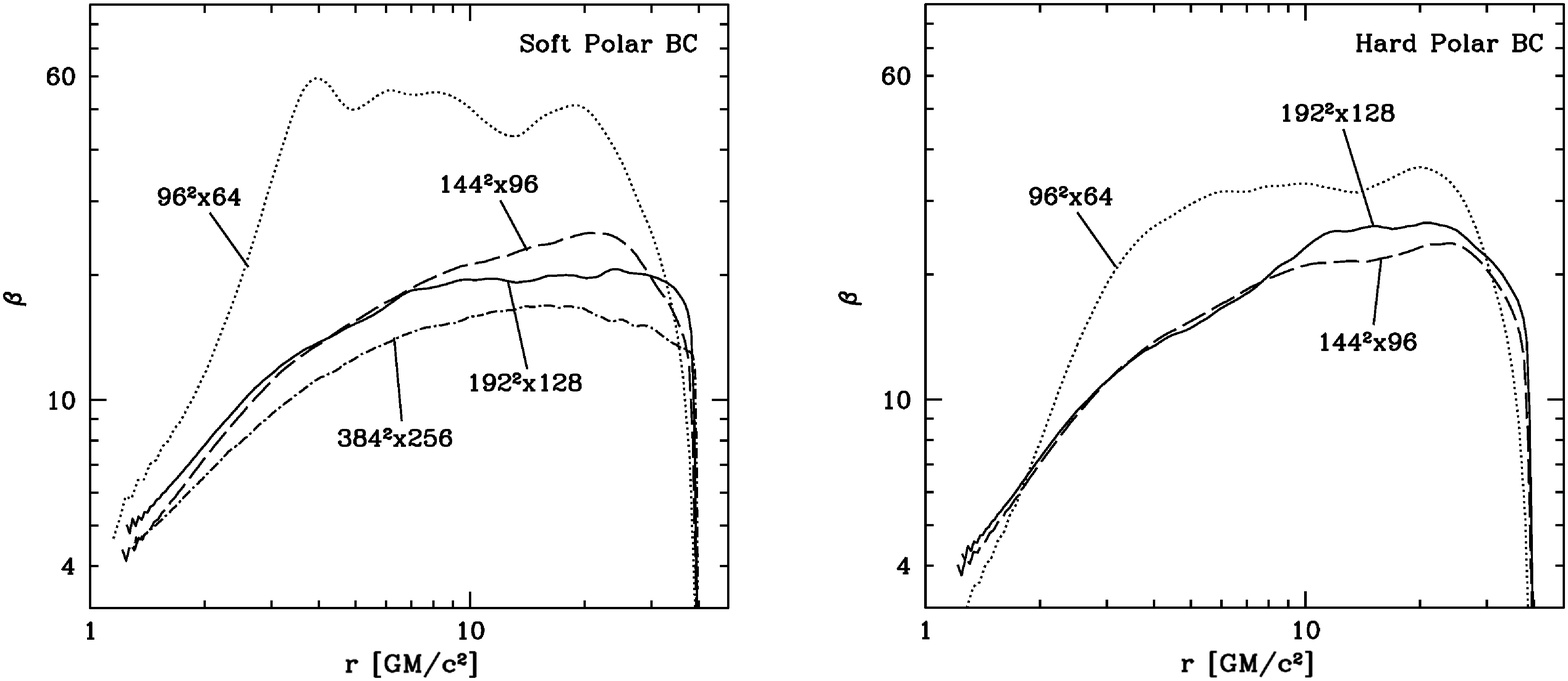}
\plotone{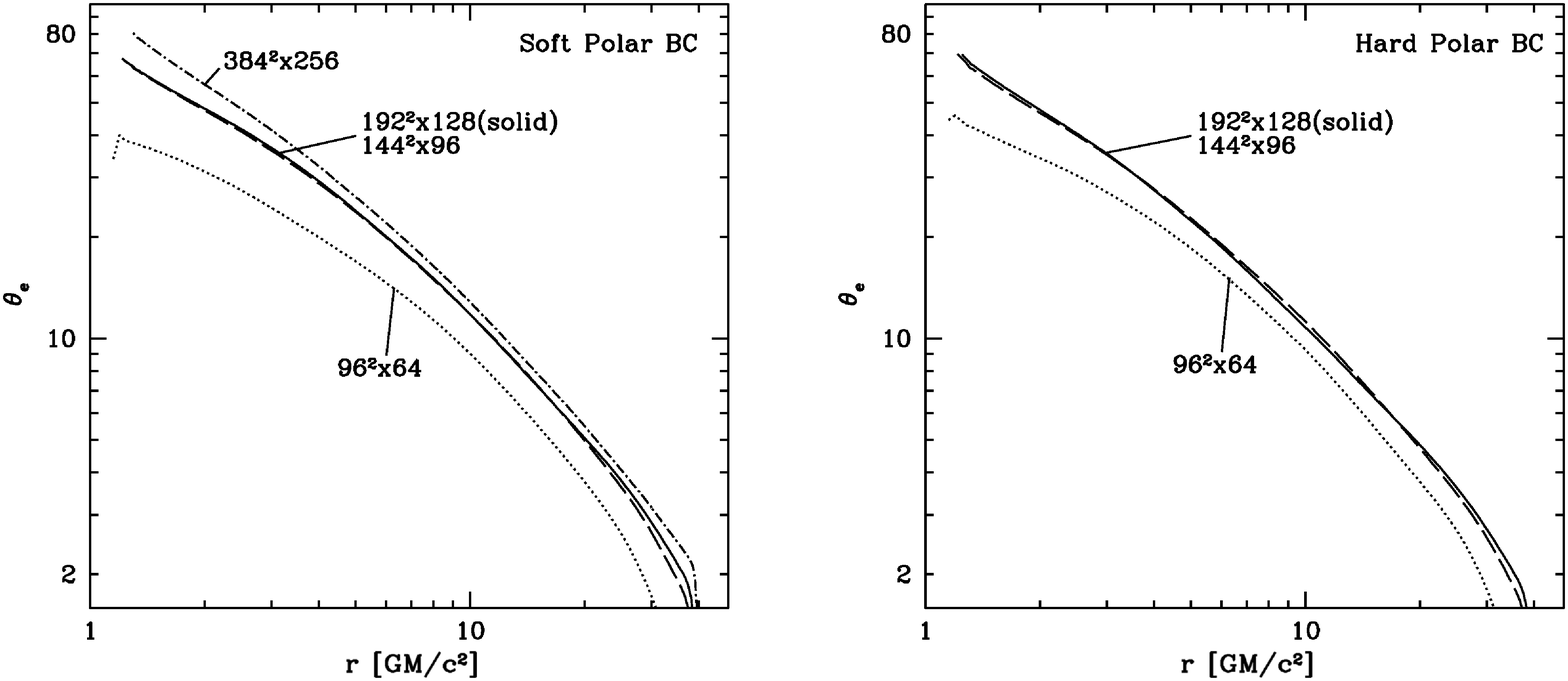}
\caption{Radial profile of plasma $\beta$ (upper row) and electron temperature $\theta_e$ (lower row) for each resolution. The columns are for the soft-polar-boundary (left) and hard-polar-boundary (right).}
\label{rprof}
\end{figure}

Figure \ref{rprof} shows the radial profile of $\beta$ and $\theta_e$
calculated using equation \ref{Ana1} for all the runs.  All time
averages run from $t = 4000$ to the end of the run; at $t = 4000$ the
disk at $r \lesssim 10$ is in a steady state  for all runs
except for the lowest resolution model, which shows a clear upward trend
in $\beta$ over the entire run.  The lowest $(96 \times 96 \times 64)$ and medium
$(144 \times 144 \times 96)$ resolution runs are averaged over 3 runs with different
initial seeds to reduce run-to-run variation.  The figure shows profiles
for both the hard and the soft polar boundary conditions described in
\S 2.  

\begin{figure}
\plotone{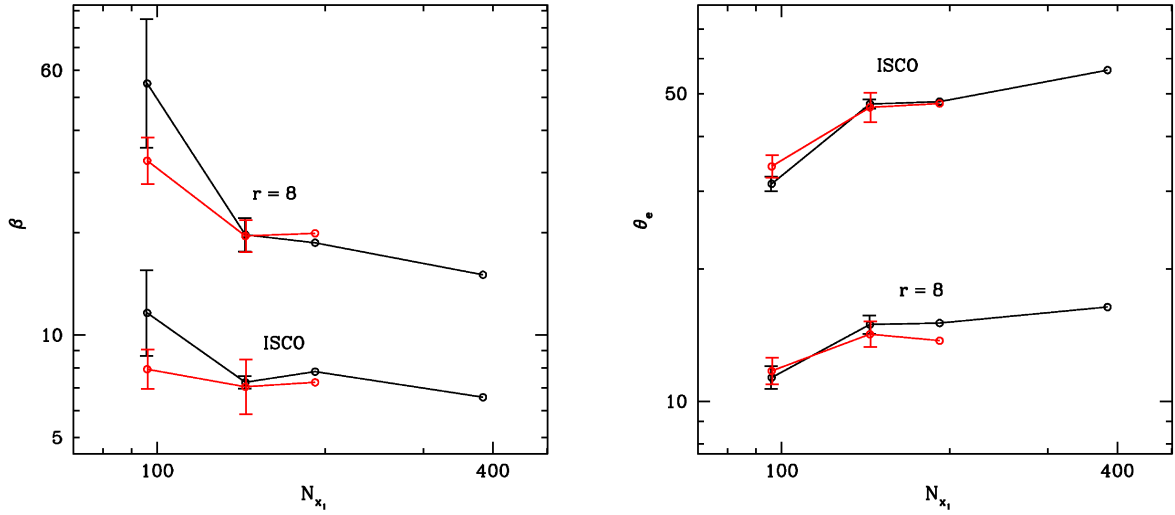}
\caption{Plasma $\beta$ (left) and electron temperature $\theta_e$ (right) plotted as a function of resolution at the ISCO ($r=2.04$) and $r=8$. The black lines are for the soft-polar-boundary and the red lines are for the hard-polar-boundary.}
\label{rprof_res}
\end{figure}

Figure \ref{rprof_res} shows $\beta$ and $\theta_e$ plotted against
radial resolution $N_{x_1}$ for $r=2.04$ (ISCO) and $8$.  The soft- and
hard-polar boundary results are shown as solid black and red lines,
respectively.  Most quantities vary sharply from $N_{x_1}=96$ to $144$ and
then far less at higher resolution.  For example, the soft polar
boundary models have $\beta(\text{ISCO}) = (11.6,7.3,7.8,6.6)$, and
$\theta_e(\text{ISCO}) = (31,47,48,57)$ at the four resolutions.

Notice that at resolutions greater than $144 \times 144 \times 96$ there are only small
quantitative differences between the hard- and soft-polar boundary
conditions, as seen in Figure \ref{rprof} and \ref{rprof_res}.  We
conclude that the effect of the polar boundary conditions on the main,
equatorial flow is small for these dimensionless variables.

What part of the variations at $N_{x_1} \ge 144$ is real variation with
resolution, and what part is run-to-run noise?  The error bars in Figure
\ref{rprof_res} show standard deviation of the three runs performed for
the lowest $(96 \times 96 \times 64)$ and medium $(144 \times 144 \times 96)$ data points with different
initial seeds.  Error bars are not available for the higher resolution
data points due to computational expense.  
The size of the error bars is comparable to the differences between models
run with different resolution.   One might hope to gain additional information by
measuring, e.g., $\beta$ at several radii and averaging the trend with resolution,
but, interestingly, the entire radial profile varies in a correlated way.   Nevertheless
Figures 2 and 3 show a clear trend of decreasing $\beta$ and $\theta_e$ with
increasing resolution.   It seems likely, therefore, that there is a genuine but
weak trend with resolution.


\section{Correlation lengths}

We have looked at one-point statistics for non-dimensional variables.
What about two-point statistics, which measure the spatial structure of
the turbulence, and in particular the correlation length?  The
correlation length is a natural measure of the outer scale of the
turbulence, and should be resolved and independent of resolution in a
converged simulation.  

We consider only the azimuthal correlation length, as this is most
straightforward to compute, and is most often under resolved in
global simulations (HGK).  The correlation function at radius
$r$ on the equatorial plane is
\begin{equation}\label{Ana5}
R(\phi) = \int_{0}^{2\pi} \delta f(\phi_0) \delta f(\phi_0 + \phi) d\phi_0 \; \text{,}
\end{equation}
where $\delta f$ is deviation from average value of variable $f$ at r.
In practice, we average $R$ in small area $r\Delta r \Delta\theta$
across the equatorial plane, normalize, and average in time:
\begin{align}
\bar{R}(r,\phi,t) &= \int_{r\pm \frac{\Delta r}{2}, \; \pm\Delta \theta} R(r,\theta,\phi,t)rdrd\theta / (r\Delta r \Delta\theta) \\
\bar{R}(r,\phi) &= \int_{t_1}^{t_2} R(r,\phi,t)/R(r,0,t) dt \; \text{.}
\end{align}
Note that the correlation function for magnetic field is defined as
\begin{equation}\label{Ana5}
R(\phi) = \int_{0}^{2\pi} \delta b^{\mu}(\phi_0) \delta b_{\mu}(\phi_0 + \phi) d\phi_0 \; \text{,}
\end{equation}
where $b^{\mu}$ is defined in \S 2.  Then 
\begin{equation}\label{Ana8}
\lambda : \bar{R}(r,\lambda) = \bar{R}(r,0)/e \; \text{.}
\end{equation}
is the correlation length at radius $r$.

\begin{figure}
\centering
\includegraphics[scale=0.3]{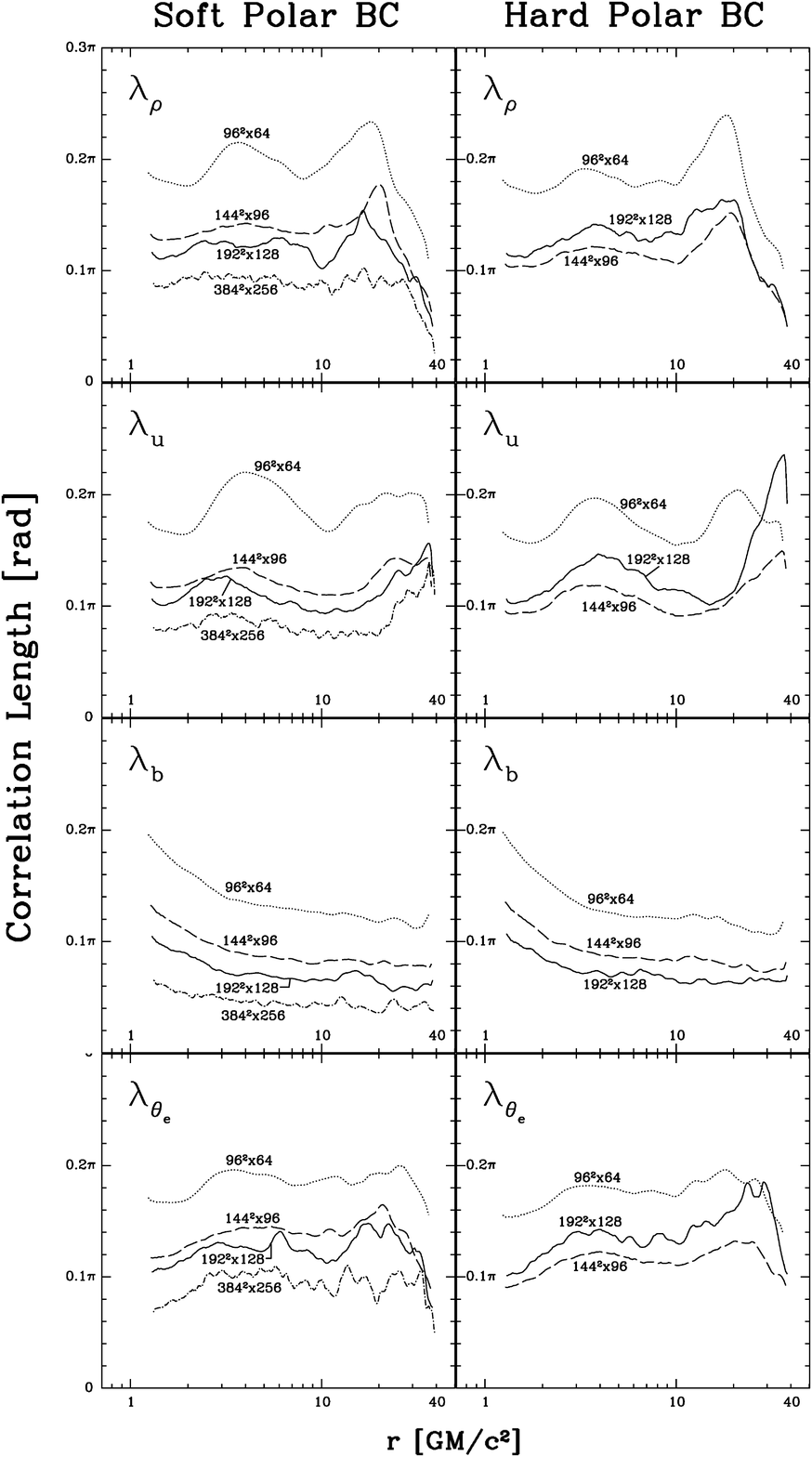}
\caption{Azimuthal correlation length as a function of radius for each resolution. From the top panel, density ($\rho$), internal energy ($u$), magnetic field ($b$), and electron temperature ($\theta_e$). The left column is for the soft-polar-boundary and right column is for the hard-polar-boundary.}
\label{correlation}
\end{figure}

Figure \ref{correlation} shows the azimuthal correlation length for
density $\rho$, internal energy $u$, magnetic field $b$, and $\theta_e$
for all runs.  Evidently the correlation lengths (angles) are nearly
independent of $r$, except close to the outer boundary where the models
are not in a steady state.  The correlation length varies between about
$0.2\pi$ at the lowest resolution to $0.1\pi$ at the highest resolution
for all variables except $b$.  Since $H/r \sim 0.3$ \footnote{The scale
height at each radius is defined as average of $\int_{\theta_0}^{\pi/2}
(\theta-\pi/2)^2\rho d\theta / \int_{\theta_0}^{\pi/2}\rho d\theta$
and $\int_{\pi/2}^{\pi-\theta_0} (\theta-\pi/2)^2\rho d\theta / \int_{\pi/2}^{\pi-\theta_0}\rho d\theta$
where $\theta_0$ is colatitude angle of the cutout $=0.017\pi$.} for
all models over a wide range in radius (Figure \ref{scale_height}), this
corresponds (assuming flat space geometry) to $1$ to $2$ vertical scale
heights.

The non-dimensional resolution $\lambda/\Delta\phi \simeq 12
(\lambda/(H/r)) (N_{x_1}/384)$ where $\Delta\phi = 2\pi/N_{x_3}$,
 is marginal even for our highest resolution simulation.  For $b$,
the correlation length of the highest resolution is smaller than that for
any other variable.  The magnetic field structure is underresolved.

\begin{figure}
\plotone{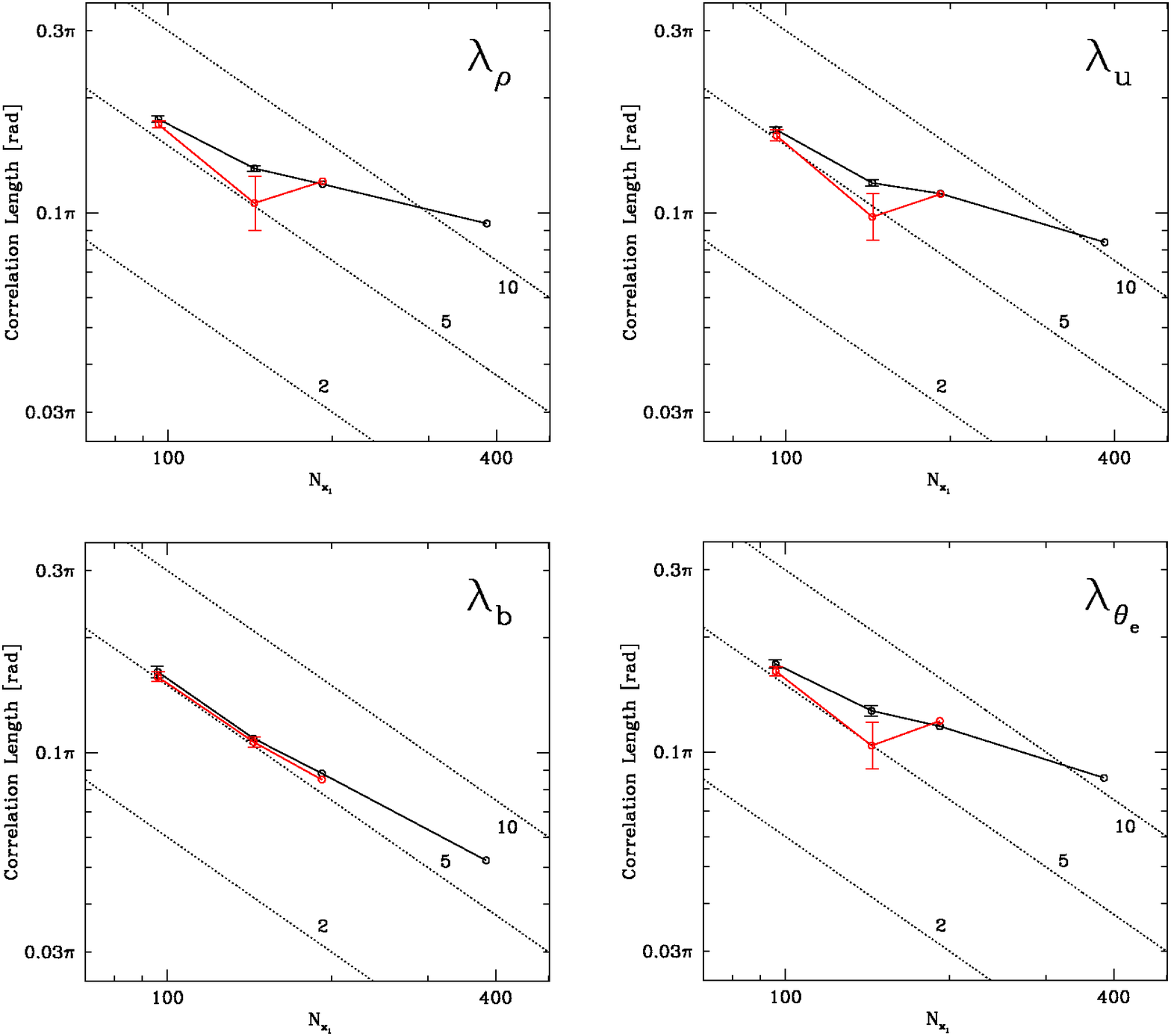}
\caption{At ISCO, azimuthal correlation length of density ($\lambda_{\rho}$), internal energy ($\lambda_{u}$), magnetic field ($\lambda_{b}$), and electron temperature ($\theta_e$) are plotted as a function of resolution. The black lines are for the soft-polar-boundary and the red lines are for the hard-polar-boundary.  Black dotted lines show a correlation length of 2, 5, and 10 grid cells,
to which correlation length size of 2, 5, and 10 grids correspond at each resolution in azimuthal direction.}
\label{correlation_res}
\end{figure}

\begin{figure}
\plotone{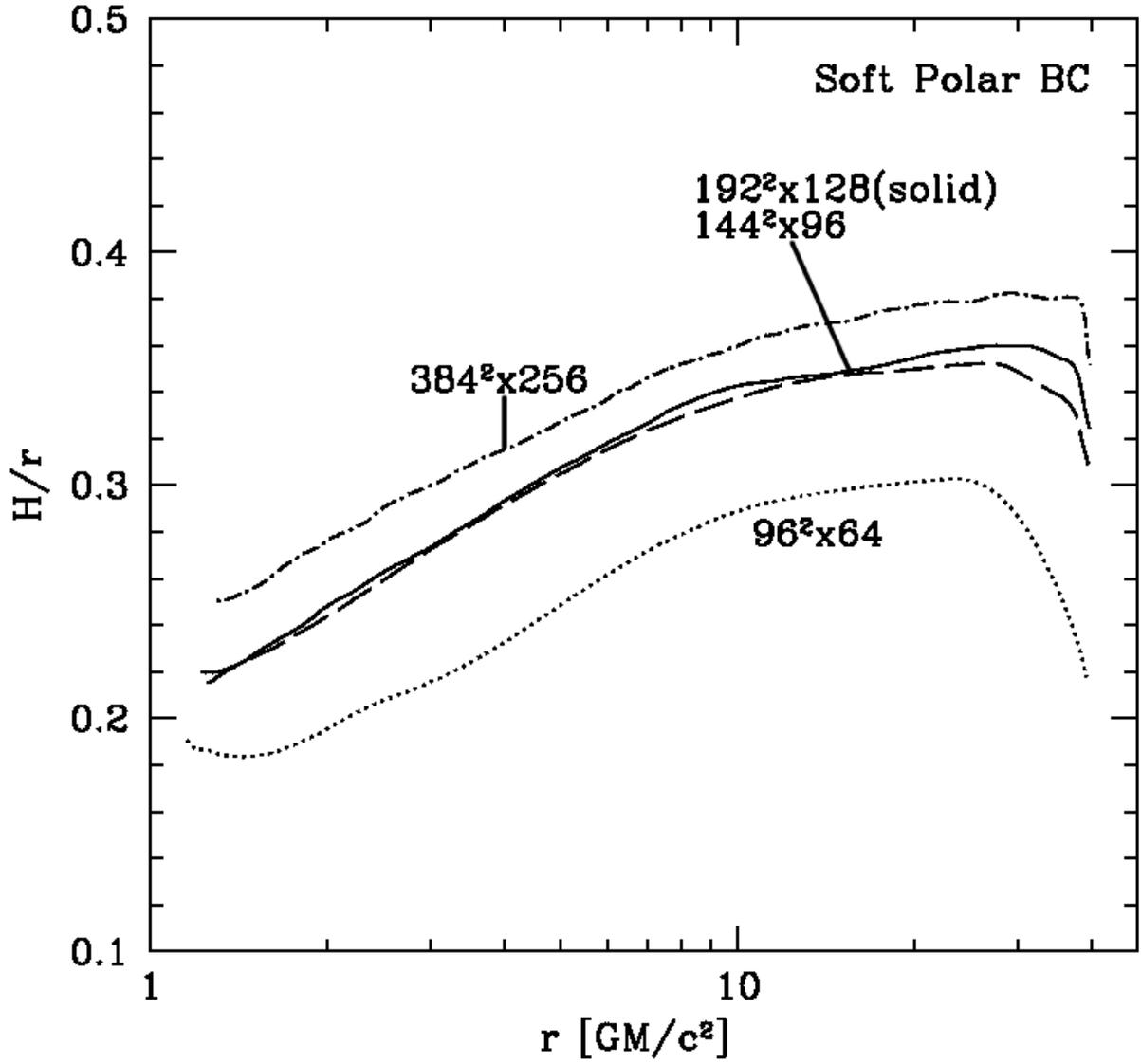}
\caption{Radial profile of the scale height $H/r$ for the runs with the soft polar boundary.
The runs with the hard polar boundary have similar profiles.
}
\label{scale_height}
\end{figure}

Figure \ref{correlation_res} plots correlation length against resolution
at the ISCO for the same variables as in Figure \ref{correlation}; here
red is the hard polar boundary and black is the soft polar boundary.
The dotted lines show how the correlation length would vary if it were
fixed at $2, 5$ and $10$ grid zones.  

For $\rho$, $u$, and $\theta_e$ (the nonmagnetic variables) the
correlation length is $\sim 5$ grid zones for the two lowest resolution
simulations.  At higher resolution-- $N_{x_1} = 192$ and $384$-- the
correlation length increases to $> 10$ grid zones, and the slope of the
change in correlation length with resolution decreases.  This suggests
that for the two highest resolution runs some structures in the
turbulence are beginning to be resolved.

For $b$, on the other hand, the correlation length decreases nearly
proportional to the grid scale, with the correlation length fixed at
around $5$ grid zones per correlation length.  There are small signs of
an increase at the highest resolution, but in light of run-to-run
variations the significance of this increase is marginal at best.  The
outer scale for the magnetic field is not resolved.

For all variables the correlation lengths for hard and soft boundary
polar conditions are consistent.  Evidently the polar boundary does not
influence the structure of turbulence in the equatorial disk.

How do these correlation lengths correspond to those found in local
model simulations?  \cite{Guan09} found in their unstratified shearing
box model that the three dimensional correlation function was a triaxial
ellipsoid elongated in the azimuthal direction and tilted into trailing
orientation.  The relationship between our azimuthal correlation length
$\lambda_b$ and the \citet{Guan09} results is
\begin{equation}
\lambda = \left( \frac{\cos^2 \theta_{tilt}}{\lambda_{maj}^2} + \frac{\sin^2\theta_{tilt}}{\lambda_{min}^2} \right)^{-1/2}
\end{equation} 
where $\theta_{tilt} \approx 15\,\rm deg$ is the tilt angle of the
correlation ellipse, and $\lambda_{maj}$, $\lambda_{min}$ are the major
and minor axis of magnetic correlation lengths.  For the best resolved
net azimuthal field model in \citet{Guan09} (y256b, which like our
global models saturates at $\beta \simeq 20$), this implies $\lambda
\simeq 0.17 H \simeq 0.05\,\rm rad$, or $0.016\pi\, \rm rad$. Therefore, it is
surprising that correlation length as large as $\simeq 0.3\, \rm rad\,\sim H$
are measured in our model for the nonmagnetic variables.

\cite{Davi10} have computed correlation lengths in stratified,
isothermal models with zero net flux.  In a model run with {\tt athena}
at a resolution of 64 zones per scale height, the implied azimuthal
correlation length (averaged over $-H < z < H$) for the magnetic field
is slightly larger than in the unstratified models of \cite{Guan09},
about $0.23 H$, or $0.02\pi\, \rm rad$.  \cite{Guan11} have also run
stratified, isothermal models at lower resolution with a {\tt zeus} type
code.  They find an implied azimuthal midplane correlation length
(similarly averaged) for the magnetic field that is even larger, about
$0.9 H$, or $0.09\pi$ radians. 
Since correlation length decreases with increasing resolution it is
possible that \cite{Guan11} are not resolving the correlation length,
and that at higher resolution the correlation length would be closer to
that measured by \cite{Davi10}.
 
The correlation length of our highest resolution run spans $0.6(H/r)$
to $0.4(H/r)$ from ISCO to $r\sim 10$ where the corresponding $\beta$
is $7$ and $16$, respectively.
This is larger than the stratified shearing box results of \citet{Davi10}
but smaller than that of \citet{Guan11}.  To resolve the correlation length
found in \cite{Davi10} we would need another factor of $2$ in linear resolution.
Note that recently \citet{Beck11} found in their global thin disk MHD simulation
that azimuthal correlation length to be about $1.3 (H/r)$ by averaging
$|z|<H$ and $5<r<11$. This is larger than our result but also falls between
\citet{Davi10} and \citet{Guan11}.


\section{Spectra}

An interesting application of GRMHD models is to simulate observations
of sources such as Sgr A* \citep{GRM, Mosc10, Hilb10, Dext09, Dext10,
Dext11}.  Are the simulated spectra converged?  

The dynamical models underlying the spectral models are run with zero
cooling, and the spectra are produced in a post-processing step.
This is self-consistent as long as the flows are advection dominated:
the accretion timescale is much shorter than the cooling timescale.  We
calculate the emergent radiation using {\tt grmonty}, a general
relativistic Monte Carlo radiative transfer code \citep{GRM}.  

{\tt grmonty} makes no symmetry assumptions and includes synchrotron emission,
absorption, and Compton scattering.  Using the rest-frame emissivity for
a hot, thermal plasma \citep{LE11} the code produces Monte Carlo
samples of the emitted photons--``superphotons" that carry a ``weight"
representing the number of photons per superphoton.  The superphotons follow
geodesics, with weight varying continuously due to synchrotron
absorption.  They also Compton scatter and produce new, scattered
superphotons with weight proportional to the scattering probability.  We
use a ``fast light'' approximation, where for each snapshot of
simulation data a spectrum is created by treating the fluid variables as
if they were time-independent.  This approximation is excellent for the
time-averaged spectra we consider here.  Superphotons that reach large
radius are collected in poloidally and azimuthally distributed bins, and
each bin produces a spectrum.  A complete description of the code is
given in \cite{GRM}.

To compare runs we generate spectra for $200-1200$ time slices
(depending on the length of the run) and time-average them.  The
spectrum of each time slice is produced from azimuthally averaged bins
that extend from $0.12\pi<|\theta-\pi/2|<0.18\pi\,\rm rad$ with respect to the
equatorial plane.

We modify the simulation-provided data in one respect before calculating
the spectrum.  The quality of the non-magnetic fluid variable
integration in the funnel region is poor due to truncation error.  In
particular the temperature can be high ($\theta_e > 10^4$) and the
particle density is determined entirely by a density floor in {\tt HARM3D}.
We therefore zero the emissivity if $b^2/\rho > 1$ to avoid
contaminating the spectrum with possibly unphysical emission.

It is necessary to fix a mass, length, and time unit to generate a
radiative model.  The combination $G M_{BH}$ sets a length and time
scale but not a mass scale because the mass of the accretion flow is
negligible in comparison to the black hole.  We set $M_{BH} = 4.5 \times
10^6 M_{\odot}$, comparable to the mass of SgrA*.  The mass unit for the
torus $\mathcal{M}$ is still free; we set it so that the $1.3\,{\rm mm}$ flux
matches the observed flux from Sgr A* of $\simeq 3.4\,{\rm Jy}$
\citep{Marr06}.  

We want to model emission from a statistically stationary accretion
flow.  Because we start with a finite mass torus and it accretes over
time, however, there is a steady decrease in density, field strength,
accretion rate, etc., as the simulation progresses.  We scale away this
long term evolution using a smooth model, as follows.  We set the mass
unit $\mathcal{M} = M_{0} s(t)$ where $M_0$ is a constant and $s(t)$ is
a two-parameter scaling function.  Then
\begin{equation}\label{Scale1}
\rho_{unit} = \mathcal{M}/(\frac{GM_{\textrm{BH}}}{c^2})^3 \quad\quad u_{unit} = \rho_{unit}c^2 \quad\quad B_{unit} = c\sqrt{4\pi\rho_{unit}} \; \text{,}
\end{equation}
or expressing with $s(t)$,
\begin{equation}\label{Scale2}
\rho_{unit} = \rho_{0}s(t) \quad\quad u_{unit} = u_{0}s(t) \quad\quad B_{unit} = B_{0}\sqrt{s(t)}
\end{equation}
where they are the unit mass density, internal energy, and magnetic
field strength, respectively, and $\rho_{0}$, $u_{0}$, and $B_{0}$ are
constants. Conversion from the simulation unit to the cgs unit is, e.g.
$\rho_{cgs}=\rho_{sim} \rho_{unit}$.


The scaling function we employ has a form
\begin{equation}\label{Scale3}
\frac{1}{s(t)} = At^{-5/3}\exp{\left(-\frac{t_{\nu}}{t}\right)}
\end{equation}
where $A$ and $t_{\nu}$ are free parameters determined by a fit to the
numerical evolution.
The form comes from fitting 1-d relativistic viscous disk models
(see Dolence et al. 2011 in prep. for more complete discussion).
Notice that without this
time-dependent scaling procedure, or with a different scaling procedure,
the spectra would vary systematically over the course of the simulation.
The spectra would also differ systematically with resolution because the
plasma $\beta$ varies with resolution.

\begin{figure}
\plotone{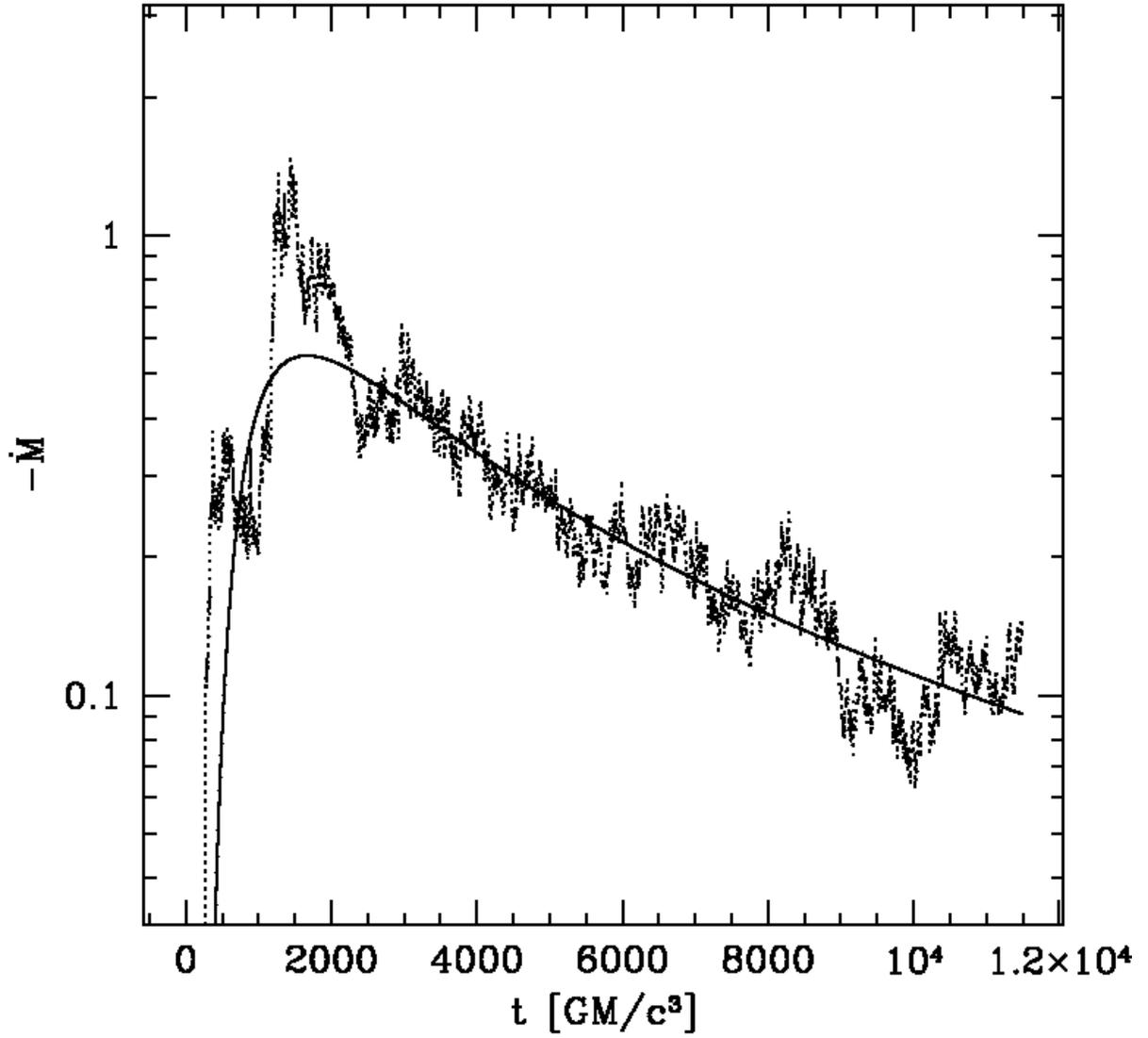} 
\caption{Time evolution of accretion rate for the run $192 \times 192 \times 128$ with hard polar boundary. Dotted line is the actual accretion rate and the solid line is a fit of the form shown in equation (\ref{Scale3}). }
\label{mdot}
\end{figure}

We fit for $A$ and the viscous timescale $t_\nu$ from simulation data
after a quasi-steady state has been reached, typically from $t = 2000$
onwards.  A sample fit to $\dot{M}$, for the $192 \times 192 \times 128$ run, is shown
in Figure \ref{mdot}. The variance of the normalized accretion rate
decreases with resolution, that is, at higher resolution the
fluctuations are smaller and equation \ref{Scale3} gives an increasingly
good fit.  The maximum of the normalized accretion rate is nearly
independent of resolution, when models with different resolution are
compared over the same time interval.

\begin{figure}
\plotone{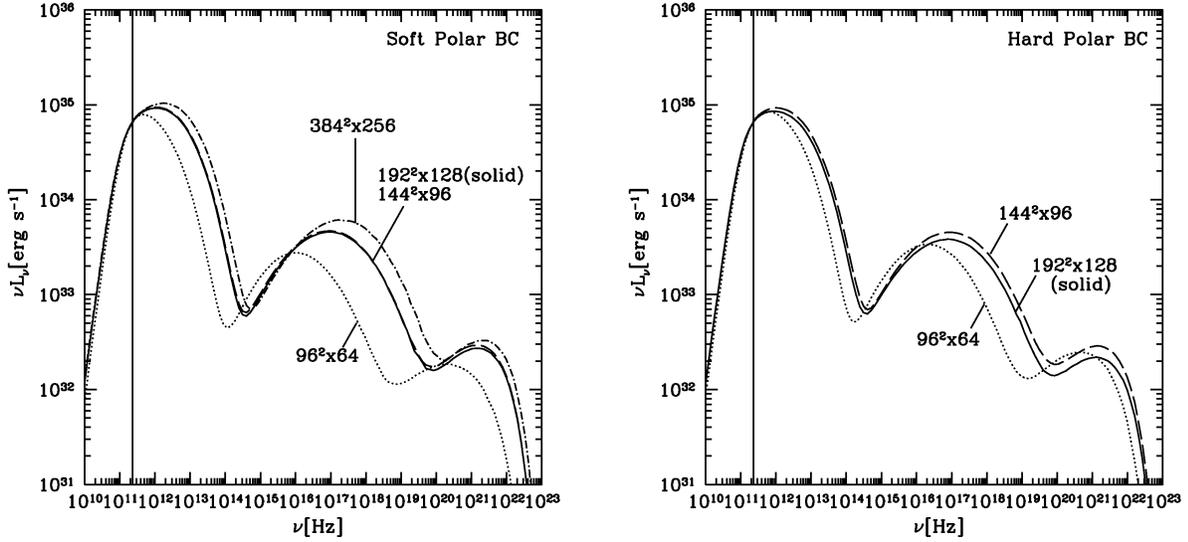}
\caption{Spectra for each resolution.  Flux is fixed to $3.4\,{\rm Jy}$ at $1.3\,{\rm mm}$ shown by the vertical solid line. The left plot is for the soft-polar-boundary and right plot is for the hard-polar-boundary. }
\label{spectra}
\end{figure}

Broadband, time-averaged synthetic spectra are shown in Figure
\ref{spectra}.  The mass unit of the torus is fixed by the condition
that $f_\nu(230\,{\rm GHz}) = 3.4\,{\rm Jy}$ for a Sgr A* model measured
at the solar circle.  The shape of the spectrum is broadly similar at
all resolutions for both polar boundary conditions.  

\begin{figure}
\plotone{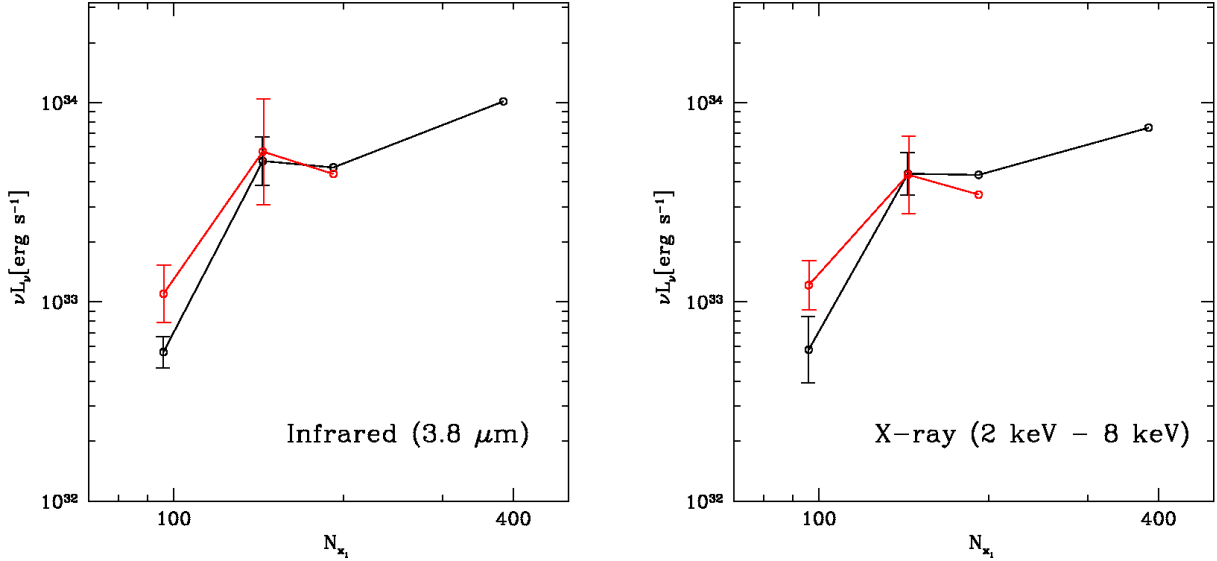}
\caption{Infrared flux density ($3.8\,\rm \mu \rm m$, left) and X-ray flux (integrated from $2 \,\rm keV$ to $8 \, \rm keV$, right) as a function of resolution. The black lines are for the soft-polar-boundary and the red lines are for the hard-polar-boundary.}
\label{spectra_res}
\end{figure}

Figure \ref{spectra_res} shows flux density plotted against resolution
in the infrared ($3.8\,\rm\mu \rm m$) and X-ray (integrated from $2\, \rm keV$ to $8\, \rm keV$)
where most of the emission is from direct synchrotron and single
Compton scatterings, respectively.
Some of the variation is likely due
to run-to-run variation, as indicated by the error bars on the $N_{x_1}
= 96$ and $N_{x_1} = 144$ models.  The flux varies with resolution by
less than about 50\% at infrared and 30\% at X-ray for $N_{x_1}>144$.  
The spectra therefore appear remarkably consistent and independent of
resolution, at least for the $M$ and $\dot{M}$ appropriate to Sgr A*.

In a sense this is not surprising, because (1) our normalization
procedure removes much of the variation that might arise from the
decrease of $\beta$ with resolution, and (2) the temperature is very
well converged.    The combined effect of the fixed flux normalization
and the variation with resolution is to strengthen the magnetic field
slightly and move the synchrotron peak slightly further into the
infrared.   This is echoed in the first Compton bump in the X-ray, which
is forced to slightly higher energy by the increase in infrared input
photons.   While we have demonstrated this for only a single set of the
model parameters ($M$, $f_\nu(230\,{\rm GHz})$), exploration of slightly
different models with similarly consistent results shows that this is
not a unique case.


\section{Summary}

We have investigated convergence of global GRMHD simulations of hot
accretion flows onto a black hole and the emergent spectrum.  We have
run GRMHD simulations for four different resolutions, $96 \times 96 \times 64$,
$144 \times 144 \times 96$, $192 \times 192 \times 128$, $384 \times 384 \times 256$ in spherical-polar coordinates.  We
have probed convergence using three diagnostics: time-averaged radial
profiles of nondimensional quantities (plasma $\beta$ and electron
temperature $\theta_e$); azimuthal correlation lengths for several
variables including the magnetic field; and artificial spectra generated
with a Monte Carlo code.  

For most of our diagnostics there are substantial differences between
the lowest $(96 \times 96 \times 64)$ and next lowest $(144 \times 144 \times 96)$ resolution, and
relative minor changes at higher resolution.  Run-to-run variations in
the lower resolution models tend to be larger than the differences
between the higher resolution $(192 \times 192 \times 128$ and $384 \times 384 \times 256)$ models.  

We find that the magnetic correlation length is {\em not converged}.  It
decreases nearly linearly with resolution, with the number of grid cells
per magnetic correlation length fixed at $\sim 5$, although we do see a
slight increase as resolution increases.  Comparison with local
model/shearing box simulations suggests that the turbulence does not
change qualitatively at higher resolution.  Such comparisons also
suggest that another factor of $\approx 2$ in linear resolution (costing
about $1.6 \times 10^7$ cpu-hours) would resolve the azimuthal magnetic
correlation length.  None of the existing simulations (local or global)
resolve scales more than a factor of $\approx 4$ smaller than the
correlation length (particularly the minor axis correlation length,
which is oriented nearly along the radial unit vector and which we have
not investigated here).   If we identify the correlation length with the
outer scale of MRI driven turbulence, as seems reasonable, then none of
these models have a resolved inertial range.

On the other hand, time-averaged synthetic spectra based on the GRMHD
models, with parameters fixed to match Sgr A*, are remarkably
reproducible from resolution to resolution.  This suggests that
simulated observations from existing simulations have some predictive
power.  We think it likely that the leading source of error in the high
resolution radiative models is now related to the underlying physical
model (particularly the fluid model treatment of the plasma, and the
absence of conduction) rather than the finite resolution of the models.

A similar convergence study has been conducted by HGK for
nonrelativistic global models.  It is worth asking whether our models
are converged according to the dimensionless resolution $Q$, the ratio
of most unstable MRI wavelength \footnote{Although $Q$ is well defined,
the background state is turbulent and there are no well defined linear
MRI modes.} to the grid cell size in the azimuthal and vertical
direction.  In the azimuthal direction, 
\begin{align}
Q_3 = & \frac{\lambda_{MRI}}{r \Delta\phi} \simeq  2 \pi \left(\frac{H}{r}\right) \frac{|B_3|}{c_s \sqrt{2 \rho}} \frac{1}{\Delta\phi}\\
\simeq & 2 \pi \left(\frac{H}{r}\right) \beta^{-1/2} \frac{1}{\Delta \phi}
\end{align}
($Q_y$ or $Q_\phi$ in HGK's notation), where $c_s \sim H\Omega$ is the sound speed.  This gives $Q_3
\gtrsim 22$ and $\gtrsim 10$ for $N_{x_1}=384$ and $192$, respectively,
for all radii less than $10$.  In the vertical direction,
\begin{equation}
Q_2 \simeq 2 \pi \left(\frac{H}{r}\right) \frac{|B_2|}{c_s \sqrt{2 \rho}} \frac{1}{\Delta\theta} = Q_3 \frac{|B_2|}{|B_3|}
\end{equation}
($Q_z$ in HGK's notation) where $\Delta\theta$ is the zone size in
Kerr-Schild coordinates at the midplane.  Since $|B_2/B_3|$ is usually $\sim 3-10$,
this gives $Q_2 \gtrsim 70$ and $\gtrsim 30$ for $N_{x_1}=384$ and $192$, respectively, for all $r <
10$.
The required $Q$ values to resolve the characteristic wavelength are
$Q_2 \gtrsim 20-60$ and $Q_3 \gtrsim 6$.
Hence, MRI in the toroidal direction is resolved but not
in the poloidal direction in these runs according to HGK's $Q$ criterion.


To summarize our findings in the form of guidance for future simulators:
(1) the resolution $96 \times 96 \times 64$ is too low.  The convergence measurements
differ by factors of several from the highest resolution runs, and the
magnetic field weakens steadily in a relative sense ($\beta$ increases)
over the course of the run; (2) the resolution $144 \times 144 \times 96$ shows early
signs of convergence except for the correlation length of the magnetic
field; (3) the resolution $192 \times 192 \times 128$ and $384 \times 384 \times 256$ differ relatively
little from each other and show signs of convergence in the azimuthal
correlation lengths, the temperature, and spectra, but {\em not} in the
correlation length of magnetic field; (4) the observed trends with
increasing resolution (to the extent that they are significant at the
highest resolution) are that $\beta$ decreases, $\theta_e$ increases,
correlation lengths decreases, and IR and X-ray fluxes increase relative
to millimeter fluxes, which we use to normalize the spectrum.


\acknowledgments

This work was supported by the National Science Foundation under grants
PHY 02-05155 and AST 07-09246, by NASA under grant NNX10AD03G, through
TeraGrid resources provided by NCSA and TACC, and by a Richard and
Margaret Romano Professorial scholarship, a NESS fellowship, NNX10AL24H, to JCD, and a University Scholar
appointment to CFG.  Part of this work was completed while CFG was a
visitor at Max-Planck-Institut f\"ur Astrophysik, and CFG would like to
thank Henk Spruit and Rashid Sunyaev for their hospitality.  The authors
are grateful to Shane Davis and to Xiaoyue Guan for providing
unpublished data from their simulations.

\newpage

\newpage

\end{document}